\documentclass{optica-article}

\journal{opticajournal}
\articletype{Research Article}
\begin{document}

\title{Ultralow Noise Microwave Synthesis Via Difference Frequency Division of a Brillouin Resonator}

\author{William Loh,\authormark{1,*} Dodd Gray,\authormark{1} Reed Irion,\authormark{1} Owen May,\authormark{1} Connor Belanger,\authormark{1} Jason Plant,\authormark{1} Paul W. Juodawlkis,\authormark{1} and Siva Yegnanarayanan\authormark{1}}

\address{\authormark{1} Lincoln Laboratory, Massachusetts Institute of Technology, 244 Wood Street, Lexington, MA 02421, USA}

\email{\authormark{*} William.loh@ll.mit.edu} %% email address is required; see note below about the corresponding author designation



\begin{abstract}
Low phase noise microwave oscillators are at the center of a multitude of applications that span the gamut of photonics and electronics. Within this space, optically-derived approaches to microwave frequency synthesis are particularly compelling owing to their unique combination of ultrawideband frequency access and the potential for resiliency to temperature and environmental perturbation via common-mode noise rejection. We demonstrate here an optical frequency divider that uses the 30 terahertz frequency gap between two stimulated Brillouin scattering (SBS) lasers as the basis for frequency division. The resulting microwave signal, centered at 10 GHz frequency, exhibits exceptionally low phase noise levels of -95 dBc/Hz and -110 dBc/Hz at 10 Hz and 100 Hz frequency offset, respectively. Moreover, the two SBS lasers, generated from a common fiber resonator, exhibits a high degree of correlated noise cancellation in their frequency difference. We measure 16.1 dB of noise rejection against intentionally applied vibrations, thus showcasing a promising pathway towards portable and robust ultralow noise photonic-microwave synthesis.
\end{abstract}

%%%%%%%%%%%%%%%%%%%%%%%%%%  body  %%%%%%%%%%%%%%%%%%%%%%%%%%
\section{Introduction}
Over the past decade, optical frequency division (OFD) has been exploited to synthesize microwave signals that are not only versatile in frequency but that also exhibit exceptional phase noise--lower than that of any microwave oscillator in existence today \cite{Fortier2011, Yao2016}. Starting from an optical frequency operating in the range of a few hundred terahertz, this signal is divided down to 10s of GHz via an optical frequency comb, undergoing a $\sim$10,000-fold reduction in both the center frequency and its associated frequency noise. Such ultrastable microwave signals, while conventionally generated by electrical means, are essential to a variety of systems including bi-static radar, synthetic aperture radar, radio astronomy, high-end analog-to-digital conversion, precision test equipment, and advanced communications networks. In addition to traditional RF electronics, low-noise microwave signals are also gaining importance in optical systems, most notably where RF signals are used to enable fine control over a pristine optical carrier through electro-optic or acousto-optic modulation. Such is the case in techniques for feedforward stabilization of ultrastable lasers \cite{Loh2020, Li2022}, for advanced manipulation of light delivery to atomic systems \cite{Nicholson2015}, and for frequency comb generation via electro-optic modulation \cite{Carlson2018, Metcalf2019}.

Previous demonstrations of OFD have so far either employed octave-spanning combs to reach division factors of >20,000 for 10 GHz microwave center frequency \cite{Fortier2011} or have generated narrowband combs from two optical carriers separated by $\sim$1 THz to achieve division factors of $\sim$100 \cite{Li2014, Li2023, Kudelin2023}. Partly due to this difference in division ratio and also partly due to the varying performance of the lasers \cite{Baynes2015} used as the basis for division, a substantial phase noise gap of >50 dB exists for OFD signals generated between these two cases. Here, we demonstrate a new scheme for OFD that uses nonlinear optics to extend beyond the 1 THz separation that previously limited the achievable division factor and thus ultimate microwave phase noise. In our approach, we employ two ultranarrow linewidth fiber SBS lasers \cite{Loh2019, Loh2020} spaced by 30 THz and bridged via the combination of electro-optic modulation and nonlinear spectral broadening, which provides an effective compromise between complexity of comb formation, phase noise performance, and common-mode noise rejection (CMNR). Importantly, this CMNR \cite {Liu2023} arises from performing OFD on the difference frequency between two SBS lasers generated from a single Brillouin resonator (difference frequency division), such that correlated instances of temperature or vibration noise are cancelled in the subtraction. We achieve phase noise levels >10 dB better than 1 THz frequency spacing OFD systems and state-of-the-art microwave oscillators, while in comparison to octave span OFD systems, we sacrifice 30 dB of phase noise performance in exchange for much reduced system complexity. In addition, we measure 16.1 dB of environmental noise rejection, which may be especially significant for improving the ruggedness and performance of future field-deployable electronic and photonic systems.

\section{Optical Difference Frequency Division System}

The general concept of our optical difference frequency divider is illustrated in Fig. 1. Initially, two SBS lasers are generated from a single SBS resonator such that vibration and temperature fluctuations are common to both lasing wavelengths (Fig. 1a). The wavelengths are chosen to be 1348 nm and 1556 nm, separated by a gap of 30 THz.
The 1556 nm SBS output is modulated by a RF signal generator whose frequency can be chosen or tuned to synthesize virtually any desired RF output, provided that the system supports modulation and photodetection at the chosen frequency (Fig. 1b). This forms a discrete set of EOM comb lines, spaced by the RF signal frequency, with each successive comb line taking on a mixture of the coherence properties of the base 1556 nm laser and multiple orders of the RF signal. A highly nonlinear fiber is used to broaden the EOM comb output such that the comb lines extend to reach the 1348 nm SBS laser (Fig. 1c). The 1348 nm SBS laser along with an adjacent comb line is photodetected to generate a heterodyne beat note that is then used to stabilize the RF output (Fig. 1d). 

The RF beat note ($f_{beat}$) at 1348 nm is represented by \begin{equation} f_{beat} = f_{SBS2} - (f_{SBS1} + m\times f_{RF}) \end{equation} where $f_{RF}$ is the driven RF modulation frequency, $f_{SBS2}$ and $f_{SBS1}$ are the frequencies of SBS lasers at 1348 nm and 1556 nm, respectively, and $m$ is an integer denoting the number of comb lines required to bridge the gap between the two SBS lasers. In our case, for a frequency separation of $\sim$30 THz and a RF modulation frequency of 10 GHz, $m$ is on the order of 3000. Thus, the comb lines here act as a lever arm that multiplies a small variation of the RF frequency to create a large frequency shift of the photodetected beat note. If the equation is instead inverted and $f_{beat}$ is locked to a reference (taken to be 0 without loss of generality), the servo then acts on $f_{RF}$ to enforce the condition 

\begin{equation} 
f_{RF} = \frac{f_{SBS2} - f_{SBS1}}{m}.
\end{equation} 

\noindent Here, the stabilized $f_{RF}$ is the desired system output representing the optical frequency difference between the two SBS lasers divided by $m$.

%\section{Figure 1 Discussion}

\begin{figure}[t b !]
\centering
\includegraphics[width = 1.02 \columnwidth]{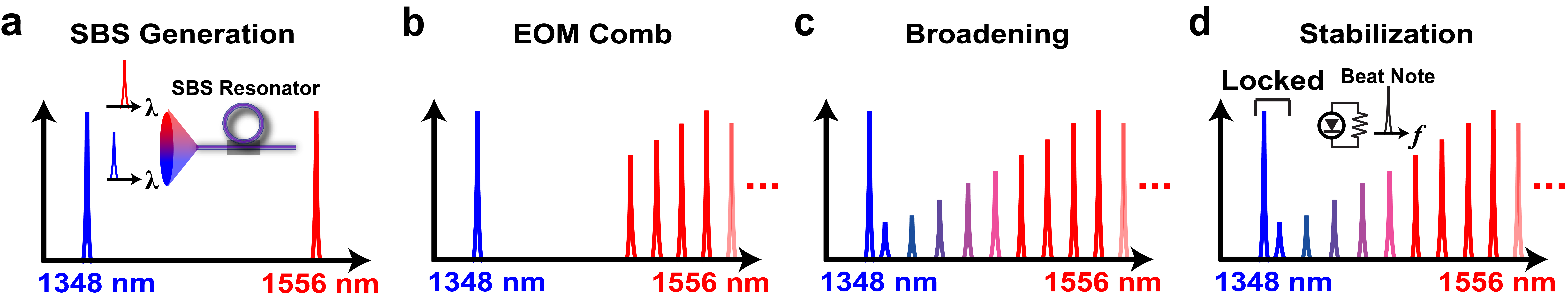}
\caption{
    \textbf{Optical Difference Frequency Division Concept.}
    \textbf{a}, Generation of two SBS lasers from a single Brillouin resonator at wavelengths of 1348 nm and 1556 nm.
    \textbf{b}, Formation of a set of comb lines around the 1556 nm SBS signal via electro-optic modulation (EOM) at a chosen RF frequency. Comb lines at longer frequencies are also present but not shown.
    \textbf{c}, Nonlinear broadening of the EOM comb to reach the second SBS signal at 1348 nm
    \textbf{d}, The beat note formed between the SBS laser and comb line at 1348 nm is locked to generate the stabilized RF output.
}
\label{fig:fig1}
\end{figure}

Such differences in optical frequency give rise to the potential for high degrees of noise cancellation if the noise is equally present on both optical frequencies. This condition is broadly applicable to instances of temperature and environmental fluctuation, especially when both optical signals are generated within a common cavity. However, since the mode numbers are different between the two optical frequencies, this cancellation will not yield identically zero. Starting from fluctuations of the cavity free spectral range (FSR), the ideal CMNR for noise is given by \begin{equation} CMNR = (\frac{\Delta f_{SBS2} - \Delta f_{SBS1}}{\Delta f_{SBS1}})^2=(\frac{m \times \Delta f_{FSR}}{n \times \Delta f_{FSR}})^2.\end{equation} $\Delta f_{SBS2}$ and $\Delta f_{SBS1}$ are frequency fluctuations of the 1348 nm and 1556 nm SBS laser arising from fluctuations of the cavity FSR (i.e., $\Delta f_{FSR}$). $n$ is the absolute mode number of the 1556 nm SBS laser such that $n \times f_{FSR} = f_{SBS1}$, and $m$ is again the mode number difference between the 1348 nm and 1556 nm SBS modes. The CMNR thus represents the ratio of the differential fluctuation between two modes relative to the absolute fluctuation of each SBS mode in isolation. For our case of two modes at 1348 nm and 1556 nm, the predicted CMNR is 16.2 dB. Note that the normalization of Eq. (3) can be to either $\Delta f_{SBS1}$ or $\Delta f_{SBS2}$, depending on the reference point.

%\section{Figure 2 Discussion}

\begin{figure}[t b !]
\centering
\includegraphics[width = 1.02 \columnwidth]{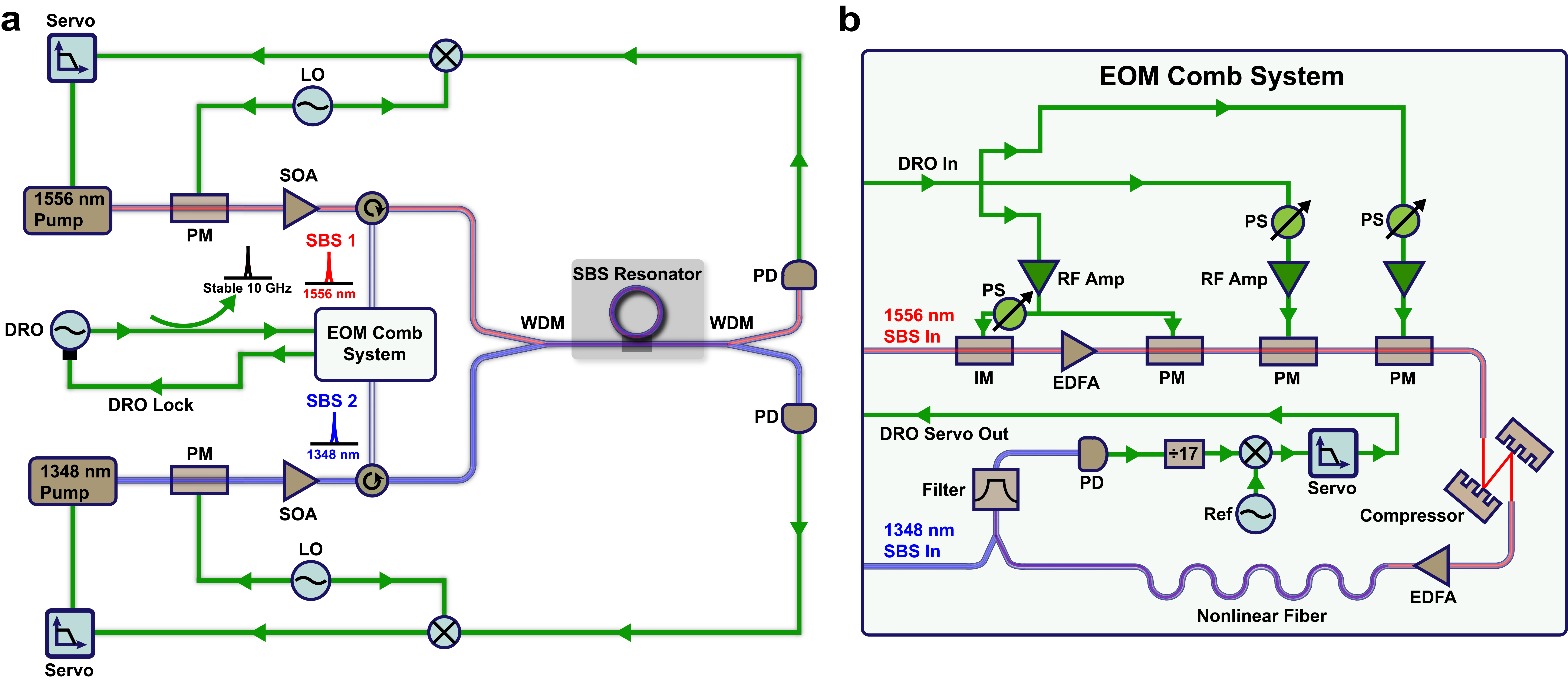}
\caption{
    \textbf{Optical Difference Frequency Division System Diagram.}
    \textbf{a}, System diagram of the SBS optical frequency divider comprising of both 1556 nm and 1348 nm SBS lasers generated from a single fiber resonator. Phase modulators (PM), semiconductor optical amplifiers (SOA), wavelength division multiplexers (WDM), photodiodes (PD), and local oscillators (LO) are used to accomplish Pound-Drever-Hall (PDH) locking of the pump lasers to the cavity resonances. The 10 GHz dielectric resonator oscillator (DRO) signal along with both 1556 nm and 1348 nm SBS outputs are sent to the EOM comb system. A portion of the DRO signal is tapped and is used as the output of the divider.
    \textbf{b}, EOM comb system diagram comprising a comb formed by intensity modulators (IM), phase modulators, and phase shifters (PS), amplified by erbium doped fiber amplifiers (EDFA) and broadened to 1348 nm in nonlinear fiber. A beat note is formed between the 1348 nm comb output and the 1348 nm SBS laser, which is used as servo feedback to stabilize the DRO frequency.
}
\label{fig:fig2}
\end{figure}

Our implementation of the optical difference frequency divider comprises two separate systems, one system for the generation of two narrow linewidth SBS lasers from a single Brillouin resonator (Fig. 2a) and a separate system to form a series of comb lines that connect the two optical frequencies (Fig. 2b). The dual wavelength SBS laser consists of two pumps centered at 1556 nm and 1348 nm and separated by 30 THz in frequency. Each pump is phase modulated and optically amplified before being sent through a circulator that extracts the SBS output from the pump input. The two pumps are then combined on a WDM and sent through a fiber Brillouin resonator having a length of 2 meters and corresponding free spectral range of 100 MHz. The resonator is encased in a copper housing and temperature controlled to keep the fiber stable against the environment. The output of the resonator is again split in wavelength via a WDM, and each pump is sent to a separate photodiode to accomplish PDH locking \cite{Drever1983} of the pump lasers to the cavity resonances. The two SBS outputs at 1556 nm and 1348 nm are extracted from the two circulators and sent to a custom Octave Photonics EOM comb system \cite{Carlson2018} along with the 10 GHz DRO output.

The EOM comb system receives the 10 GHz DRO output and splits the RF signal three ways. Each path is RF amplified and sent through independent phase shifters before being used to drive 4 modulators (1 intensity modulator and 3 phase modulators) that act on the 1556 nm SBS output to form a frequency comb. The phase shifters ensure proper phasing such that all modulators coordinate in unison to form a coherent frequency comb. The output is sent through a grating compressor and an erbium doped fiber amplifier before being broadened by a nonlinear fiber to reach 1348 nm. The broadened comb output and the 1348 nm SBS laser output are combined and filtered to extract only a 1 nm span centered around 1348 nm. The output is photodetected to generate a comb-SBS beat note that is divided by 17 and mixed with a reference oscillator to form a phase locked loop that stabilizes the beat note to the reference oscillator. The servo feedback signal is sent back to actuate on the DRO frequency, and a portion of the stabilized DRO is tapped to create the system output.

\section{Results and Measurement}

Through a combination of modulation and nonlinear broadening, our EOM comb system bridges the 30 THz gap to connect the SBS lasers at 1348 nm and 1556 nm. Figure 3a shows the optical spectrum of the base laser with three successive stages of modulation. With just intensity modulation and one stage of phase modulation, the 1556 nm SBS reduces in signal height and redistributes this energy to the wings for a total spectral width of 3.7 nm. After employing an additional stage of phase modulation, the spectral width increases to 6.4 nm. Finally, after adding the third stage of phase modulation, the EOM comb reaches its eventual span of 8.9 nm. To further extend the EOM comb to 1348 nm, the comb output is compressed, amplified, and broadened using highly nonlinear fiber. Figure 3b shows the final comb spectrum that starts at 1556 nm and reaches beyond 1300 nm wavelength. Accounting for losses and the 5 nm resolution bandwidth, the power per comb line at the output of the broadening stage is 2.4 $\mu$W at 1348 nm.

The mechanism for OFD in our system critically relies on an effect akin to a lever arm where a small variation of the base 10 GHz DRO frequency builds to a large cascaded shift of the beat note at 1348 nm. This cascading of RF frequency fluctuations with each successive comb line creates a large multiplication factor for RF frequency, which when inverted by locking the beat note to a constant, creates the desired means for OFD. Figure 3c depicts measurements of the beat note frequency as the base DRO oscillator is intentionally varied in frequency around 10 GHz. For step sizes of 30 kHz in the DRO frequency, the corresponding shift in the beat note frequency is $\sim$88.5 MHz. Figure 3d plots the measured correspondence between shifts of the DRO and beat note frequencies. The extrapolated slope is 2960, which matches the frequency multiplication factor expected for the $\sim$30 THz gap between 1348 nm and 1556 nm subdivided by 10 GHz spaced comb lines.

%\section{Figure 3 Discussion}

\begin{figure}[t b !]
\centering
\includegraphics[width = 1 \columnwidth]{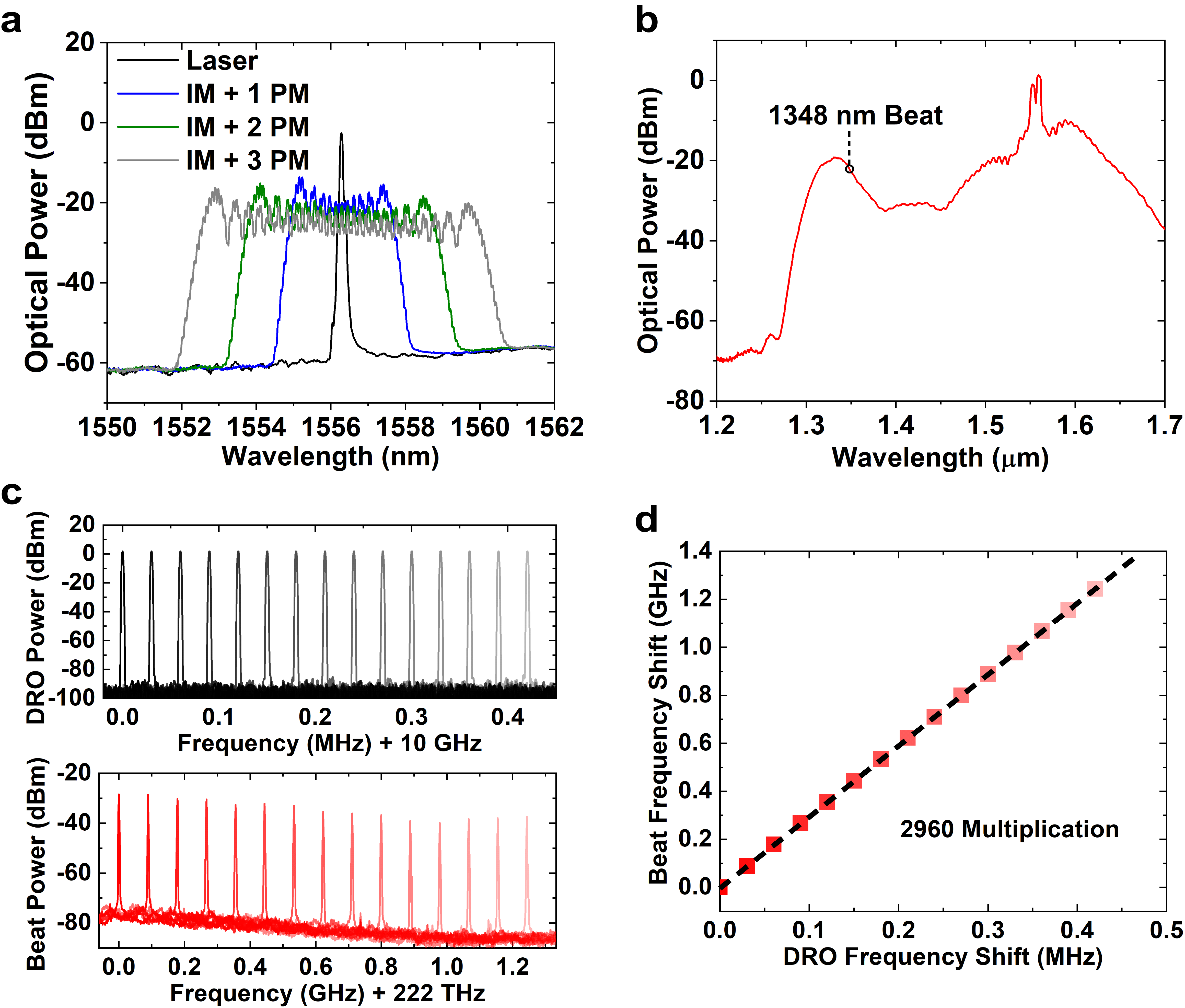}
\caption{
    \textbf{EOM Comb Measurements}
    \textbf{a}, EOM Comb spectrum with successive stages of modulation. The spectrum was taken with 60 pm resolution bandwidth and includes 12.3 dB of combined loss through various taps and component loss.
    \textbf{b}, Comb output after compression and highly nonlinear fiber broadening. The resolution bandwidth used was 5 nm, and the total optical loss through tap ports was 15 dB.
    \textbf{c}, Measurement of the lever arm effect starting from applied discrete shifts of the DRO frequency to large frequency deviations of the 1348 nm beat note.
    \textbf{d}, Extrapolated frequency multiplication factor of the EOM comb determined by linear fit.
}
\label{fig:fig3}
\end{figure}

The use of optical frequency difference as the basis for division offers additional means for noise suppression, if such instances of noise are highly correlated to both optical frequencies. To demonstrate this, we intentionally apply a sinusoidal vibration to the optical platform that which the SBS laser system rests on. The frequency of this vibration is swept from 10 Hz to 1 kHz, with only the portion between 10 Hz and 100 Hz plotted in Fig. 4a for purposes of clarity. The phase noise is measured for two separate cases--when the 1348 nm and 1556 nm SBS lasers are generated from a single common cavity and when the two lasers are generated from separate cavities. In the separate cavity case, the applied vibration is solely to the 1556 nm cavity. The ratio of the phase noise measured between the two cases yields the CMNR of our OFD system. Figure 4b depicts the measured CMNR from 10 Hz to 1 kHz. The average CMNR value across all measurements is 16.1 dB. This agrees well with the predicted CMNR of 16.2 dB based on the 30 THz spacing between the SBS lasers.

%\section{Figure 4 Discussion}

\begin{figure}[t b !]
\centering
\includegraphics[width = 1 \columnwidth]{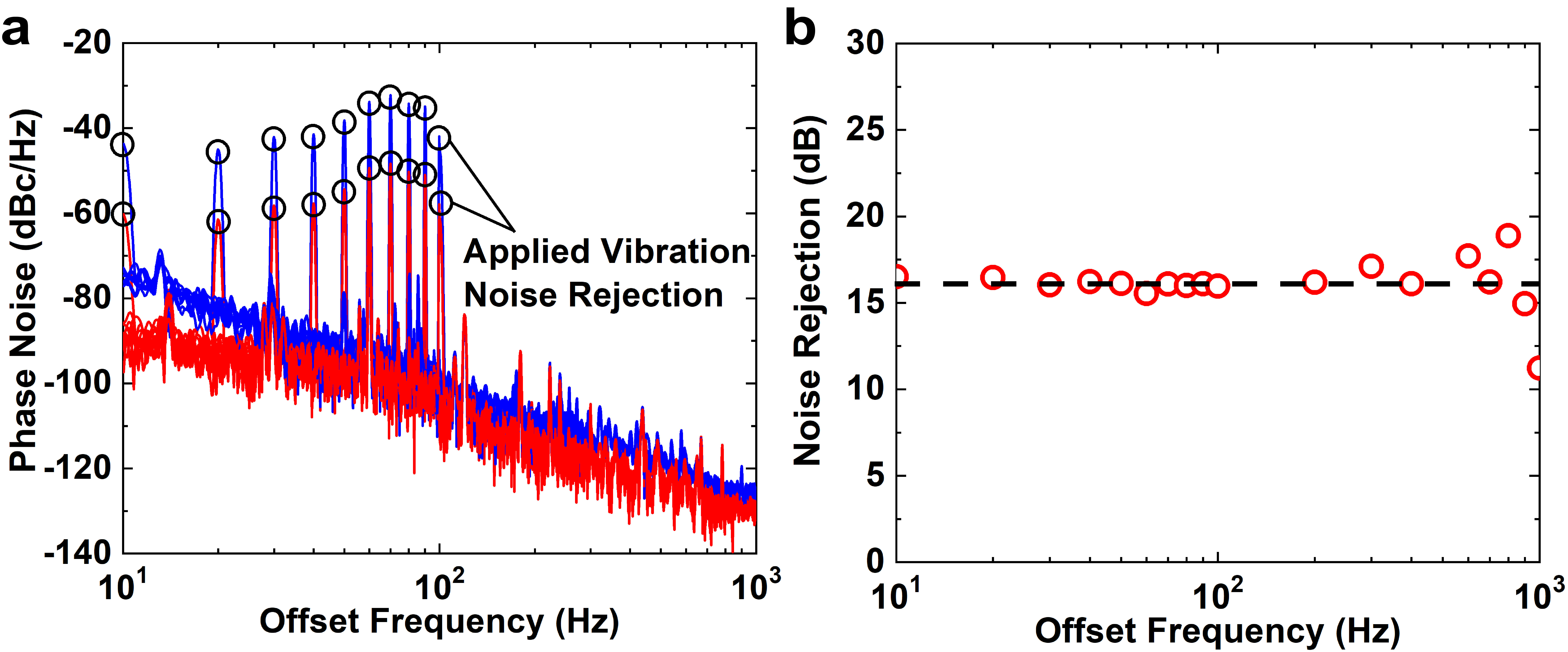}
\caption{
    \textbf{Measurements of CMNR for Applied Vibration.}
    \textbf{a}, Phase noise degradation for applied sinusoidal vibration at frequencies ranging from 10 Hz to 100 Hz in intervals of 10 Hz. The phase noise is measured for cases when the 1348 nm and 1556 nm SBS lasers share a common cavity (red solid line) and when they are generated from separate cavities (blue solid line).
    \textbf{b}, Measured CMNR for applied sinusoidal vibration at frequencies between 10 Hz and 1 kHz. The dashed line indicates the numerical average over all CMNR measurements.
}
\label{fig:fig4}
\end{figure}

We characterize the performance of our optical difference frequency divider by measuring the spectral purity of the 10 GHz output before and after stabilization. The RF spectrum is shown in Fig. 5a over a span of 1 kHz. When locked, the output narrows to a near delta-function peak, limited by the resolution bandwidth of 1 Hz. However, when the system is unlocked, the output reverts to the original DRO noise and broadens significantly. Nearly 60 dB of noise differential is observed for offset frequencies of 100--500 Hz from the carrier.

Figure 5b compares the measured phase noise of our divider output when the 1348 nm and 1556 nm SBS lasers are generated from a common cavity and when they are generated in separate cavities. This is directly analogous to the CMNR investigations of Fig. 4 except with no intentional induced vibration. As much of the SBS laser noise is limited by thermorefractive noise \cite{Gorodetsky2004, Notcutt2006, Matsko2007, Webster2008, Sun2017, Lim2019, Huang2019, Panuski2020, Jin2021}, the phase-noise is largely cancelled in the optical frequency difference. As a result, the common-cavity divided output is lower in noise, particularly at lower offset frequencies, compared to the separate cavity configuration. This also provides some indication that at higher offset frequencies beyond 1 kHz, where the noise of the common and separate cavity arrangements become indistinguishable, that some other noise mechanism becomes the limitation.

%\section{Figure 5 Discussion}

\begin{figure}[t b !]
\centering
\includegraphics[width = 1 \columnwidth]{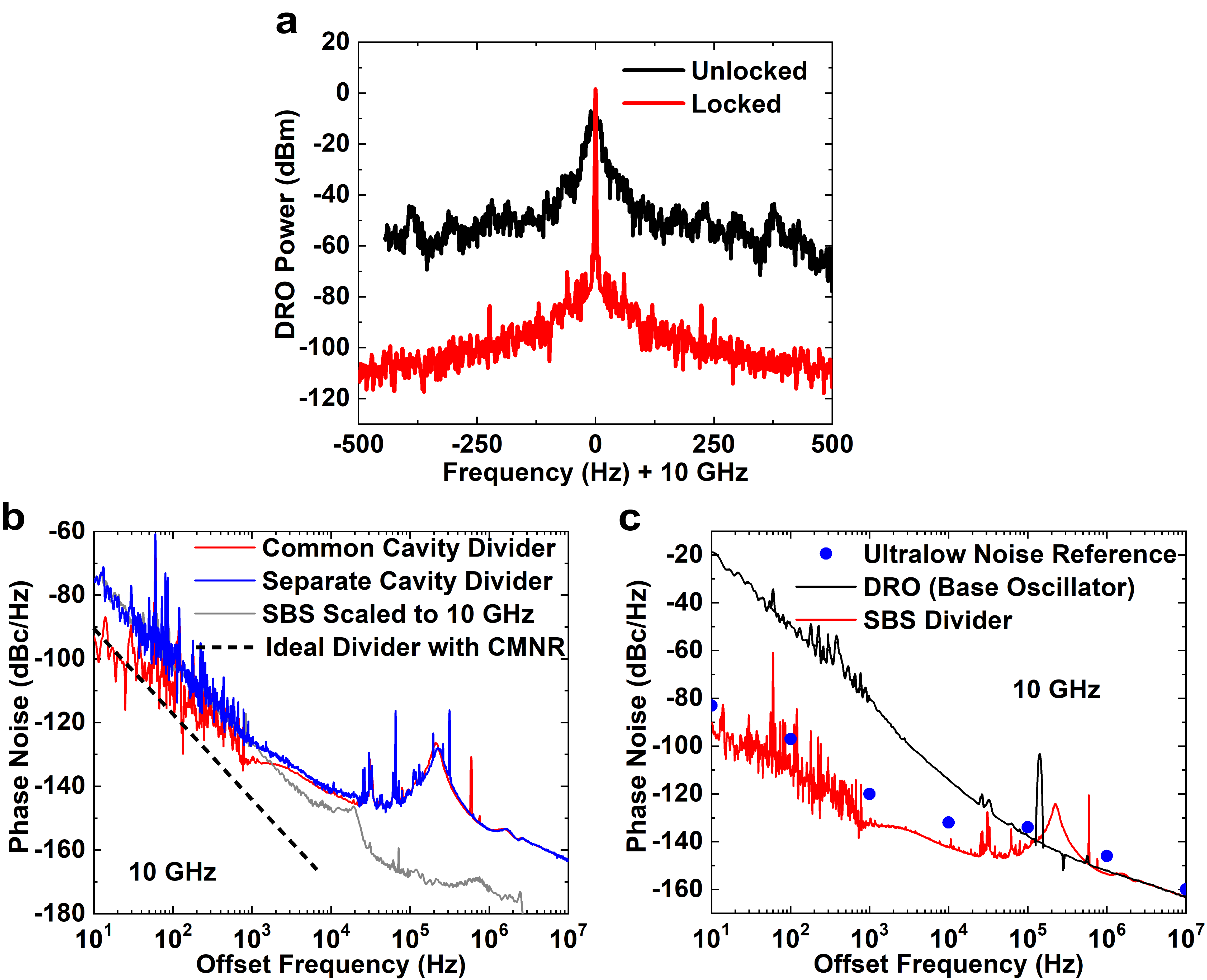}
\caption{
    \textbf{Measurements of Optical Difference Frequency Divider Noise.}
    \textbf{a}, RF Spectrum of 10 GHz DRO output before (black solid line) and after (red solid line) servo lock. The resolution bandwidth used was 1 Hz.
    \textbf{b}, 10 GHz Phase noise of the optical difference frequency divider using a common SBS laser cavity (red solid line) and separate 1348 nm and 1556 nm cavities (blue solid line). The phase noise of an SBS laser numerically divided to 10 GHz (gray solid line) is also shown for comparison. The dashed line indicates the projected close-in phase noise with ideal CMNR.
    \textbf{c}, Phase noise comparison between SBS optical difference frequency divider (red solid line), base 10 GHz DRO (black solid line), and a high-performance Rohde $\&$ Schwarz reference oscillator with ultralow phase noise option (Rohde $\&$ Schwarz SMB100B-B711, blue dots).
}
\label{fig:fig5}
\end{figure}

We investigate this noise by numerically dividing the measured SBS laser's noise at optical frequencies \cite{Loh2019, Loh2020} down to 10 GHz. For this calculation, we assume the SBS lasers at 1556 nm and 1348 nm to be generated by separate cavities but otherwise similar in noise performance. Thus, the optical difference yields twice the noise spectral density of each laser alone. We divide this spectral density by the square of the division factor ($m=2960$) and plot the resulting phase noise for comparison. The 10 GHz scaled SBS trace follows closely the separate cavity arrangement at low offset frequencies, as the cavities were assumed independent in the division calculation. However, beyond 1 kHz offset frequency, the measured noise of both the common and separate cavity configurations rise above that of the calculated 10 GHz scaled SBS laser. This suggests that rather than the noise becoming less correlated between the 1348 nm and 1556 nm SBS lasers at these offset frequencies, the noise is instead limited by an unrelated external source that only affects one of the two lasers. One possible such noise source is the EDFA, which solely amplifies the 1556 nm light prior to spectral broadening. The estimated performance of our SBS divider under conditions of ideal division and CMNR is plotted for reference. We observe that the measured common cavity SBS noise to follow this ideal case at low offset frequencies below 100 Hz. Despite the fact that our measured noise deviates at higher offset frequencies, the ideal division and CMNR case nevertheless showcases the potential performance once the system noise is minimized.

Figure 5c compares the 10 GHz SBS optical difference frequency divider to some commercial options for signal generation at 10 GHz. The base DRO oscillator is observed to be many orders of magnitude worse than the performance of the SBS divider ($\sim$75 dB at 10 Hz offset), which illustrates the power of using optical techniques for microwave frequency synthesis. The SBS divider reaches -95 dBc/Hz and -110 dBc/Hz at 10 Hz and 100 Hz frequency offsets, and its noise performance reverts to the DRO noise past the locking bandwidth (200 kHz). In comparison to a state-of-the-art Rohde $\&$ Schwarz synthesizer, the SBS divider phase noise is improved by $\sim$10 dB.

\section{Conclusion}

Optical techniques are compelling as a way to synthesize microwave signals with properties typically inaccessible to purely electrical means alone. Here, we demonstrated an optical difference frequency divider that divides an SBS laser down to an ultralow phase noise 10 GHz signal--and that can be extended to any frequency the electro-optic modulators and photodetectors can provide response to. In addition to microwave frequency synthesis, the system output doubles as an optically-stabilized comb, whose stability in terms of center frequency and repetition rate is directly tied to that of the base lasers used. By taking differences in optical frequency, correlated noise that appear on both frequencies becomes suppressed in the subtraction--leading to substantial noise reduction. With the potential for scalability via chip integration \cite{Zhang2019, Loh2022}, this method of optical difference frequency division offers a robust, compact, and higher-performing alternative to the currently dominant crystal oscillator solutions in existence today.

\section{Disclosures}

The authors declare no conflicts of interest.

\section{Acknowledgments}

We thank David Carlson, Daniel Hickstein, and Zachary Newman for their help with setting up the electro-optic frequency comb. This work was sponsored by the Under Secretary of Defense for Research and Engineering under Air Force contract number FA8721-05-C-0002. Opinions, interpretations, conclusions and recommendations are those of the authors and are not necessarily endorsed by the US Government.

%%%%%%%%%%%%%%%%%%%%%%% References %%%%%%%%%%%%%%%%%%%%%%%%%

\bibliography{SBS_Frequency_Division}

\end{document}